\documentclass[12pt]{iopart}
\input {epsf.sty}
\usepackage{epsfig}
\begin{document}

\title{Magnetic Impurities in the Pnictide Superconductor  Ba$_{1-x}$K$_{x}$Fe$_{2}$As$_{2}$}

\author{Sutirtha Mukhopadhyay$^1$, Sangwon Oh$^1$, A M Mounce$^1$, Moohee Lee$^2$, W P Halperin$^1$, N Ni$^3$, S L Bud\'{}ko$^3$, P C Canfield$^3$, A P Reyes$^4$ and P L Kuhns$^4$ }
\address{$^1$Department of Physics and Astronomy, Northwestern University, Evanston, Illinois 60208, USA.\\$^2$Department of Physics, Konkuk University, Seoul 143-701, South Korea\\$^3$Ames Laboratory US DOE and Department of Physics and Astronomy, Ames, IA 50011, USA\\$^4$National High Magnetic Field Laboratory, Tallahassee, FL 32310, USA.} 
\ead{w-halperin@northwestern.edu}
\date{Version \today}

\begin{abstract} NMR measurements have been performed on single crystals of Ba$_{1-x}$K$_{x}$Fe$_2$As$_2$ (x = 0, 0.45) and
CaFe$_2$As$_2$ grown  from Sn flux.  The Ba-based pnictide crystals contain significant amounts of Sn in
their structure,
$\sim 1$\%, giving rise to magnetic impurity effects evident in the  NMR spectrum and in the magnetization.  Our
experiments show that the large impurity  magnetization is broadly distributed on a microscopic scale, generating
substantial magnetic field gradients.  There is a concomitant
$20$\% reduction in the transition temperature which is most likely  due to magnetic electron scattering. We
suggest that the relative robustness of superconductivity ($x=0.45$) in  the presence of severe magnetic
inhomogeneity might be accounted for by strong spatial correlations between impurities on the coherence length scale.
\end{abstract}


\maketitle

\section{Introduction}

The new FeAs based superconductors are interesting candidates for the  study of competing magnetic and superconducting
order. After the discovery of superconductivity~\cite{Kamihara2008, Takahashi2008} in electron-doped pnictides, 
their hole-doped counterparts were  synthesized~\cite{rotter2008, ni2008} with K doped BaFe$_2$As$_2$, Na
doped  CaFe$_2$As$_2$~\cite{wu2008}, as well as K and Cs doped SrFe$_2$As$_2$~\cite{GFchen2008, sasmal2008}.  The
parent compounds  for both the electron and hole-doped materials undergo structural phase transitions from tetragonal
to  orthorhombic on cooling, accompanied by an antiferromagnetic spin density wave (SDW) state.   When the parent 
compounds are doped both the structural transition and the SDW are suppressed and superconductivity appears. 

In fact, there is growing evidence for {\it coexistence} of  static  magnetism and superconductivity extending into the
superconducting region of the FeAs based  superconductors. Muon spin relaxation ($\mu$SR) experiments~\cite{aczel2008,
drew2008, goko2008} in both  electron and hole-doped pnictides have revealed substantial magnetism in the
superconducting state. A recent  report on neutron diffraction in Ba$_{1-x}$K$_{x}$Fe$_2$As$_2$ 
 showed  that magnetism and superconductivity may coexist even up to 40\% of K doping~\cite{Hchen2008}. In order
to  understand the relationship between magnetism and superconductivity in the FeAs superconductors it is of  central
importance to determine if it is magnetic order or magnetic disorder that competes with  superconductivity; whether the
magnetism is local or exists in a separate phase, or is possibly isolated in various regions of the  sample. This is
the main thrust of our work where we use NMR as a microscopic  probe of magnetic impurities.

It is well established~\cite{Abr61} that $s$-wave superconductivity is  suppressed by magnetic disorder where
quasiparticle scattering from magnetic impurities is an  effective pair breaking process, often called spin-flip
scattering.  In contrast, both potential and  magnetic scattering in non-$s$-wave superconductors are effective at
breaking Cooper pairs~\cite{Lar65}. Consequently, the study of the robustness of the superconducting state in the
presence of magnetic  impurities, compared to non-magnetic impurities, can be taken as one indication for the symmetry
of the  superconducting order parameter. Additionally, impurities can play a key role in the pinning of flux  which may
be essential in achieving high critical current densities in some applications.

We have investigated the effect of  magnetic impurities in the normal  state of crystals of the superconducting
compound Ba$_{0.55}$K$_{0.45}$Fe$_2$As$_2$, which we will refer to  henceforth simply as BaK122, using
$^{75}$As NMR (gyromagnetic ratio,  $\gamma=7.2919$ MHz/T (used as the reference for calculating the Knight shift in
this work); spin, $I  =3/2$;  and natural abundance, 100\%). We compare some of our results with undoped crystals,
Ba122, and  Ca122.  The crystals were grown and characterized by Ni
{\it et al.}\cite{ni2008}.  The use of a Sn flux~\cite{ni2008} leads to
$\approx 1$\% Sn impurity, likely  on the As site, for Ba122 and BaK122.  In contrast the Ca122 crystals have
negligible amounts of Sn in the structure~\cite{ni2008a}. The potassium doped crystals become superconducting with a transition
temperature of
$\sim 30$ K.  By comparison with the pure compound~\cite{rotter2008}  with the same doping we can infer that the
$T_c$ suppression is
$\sim 8$ K.   The principal question we address is how uniformly  distributed are these impurities and are they coupled
to the conduction electrons.  We have found that they  are strongly magnetic; that they are locally distributed on a
microscopic scale; and that they appear to  be closely coupled to the conduction electrons. It is unusual that a
compound remains superconducting in  the presence of such significant amounts of magnetism.

Impurities play an important role in modifying the transition  temperature and the amplitude of the order parameter in
unconventional pairing systems and have been studied  extensively in the case of
$^3$He superfluid\cite{Hal08} and reviewed by Balatsky {\it et  al.}~\cite{Bal06} for superconductors.  In high-$T_c$
cuprates  NMR measurements of $^{17}$O in the Cu--O plane,  and of the impurity ion itself, have been
performed~\cite{Alloul2009}.  The cation substitution for planar copper of  Zn~\cite{alloul1991},
Ni~\cite{bobroff1997}, Li~\cite{bobroff1999} in YBCO, and Al  substitution in
LSCO~\cite{ishida1996} suppress
$T_c$ and create local moments.  The fractional reduction in the  transition temperature,
$\Delta T_c$/$T_c$  per \% impurity (in the low concentration limit) for these
examples~\cite{alloul1991,bobroff1997,bobroff1999,ishida1996, Xia87,  Chi91} is, 0.036 (Ni), 0.12 (Zn), 0.058(Li) and
0.5 (Al) which are similar to our sample of  BaK122 with
$\sim$ 0.2 (Sn) assuming
$\sim$ 1\% of Sn substituted into the structure.  On the other hand,  the magnetization of the BaK122 is almost two
orders of magnitude larger than that of a LSCO sample with  3\% impurity of Al for which superconductivity is
completely eliminated.

Chemical substitution in the 122 family has a different role as  compared with the cuprates in that superconductivity
only appears by inhibiting the structural  transition as can be achieved by insertion of Co into
Ba122~\cite{sefat2008}, Sr122~\cite{leithejasper2008}, and Ni  into
Ba122~\cite{li2008}.  However, the nature and effect of impurity scattering on the superconducting  state
in the pnictides is not yet well-established.  M\"{o}ssbauer spectroscopy has  revealed~\cite{nowik2008} the presence
of possible magnetic phases of FeAs, FeAs$_2$,  and Fe$_2$As in a significant amount. A zero  field NMR spectrum of
$^{75}$As at $\sim 265$ MHz in Sm oxy-pnictides~\cite{sidorenko2008}  was attributed to the possible existence of some
magnetic impurity phase. Although the $\mu$SR  experiments~\cite{aczel2008, drew2008, goko2008} have identified static
magnetism in both pnictides and  oxy-pnictides, they could not ascertain its origin. A $\mu$SR
measurement~\cite{baker2008} on  FeAs, and  FeAs$_2$ seems to rule out the possibility of these two impurity phases as
being responsible for the  observed magnetism. In this paper, we report normal-state NMR measurements on crystals
containing $\sim  1$\% Sn impurity of superconducting Ba$_{0.55}$K$_{0.45}$Fe$_2$As$_2$ and non-superconducting 
BaFe$_2$As$_2$ and CaFe$_2$As$_2$.

\section{Experimental Methods} We have performed  $^{75}$As NMR  experiments from 6.4 to 14 T at Northwestern
University and the National High Magnetic Field Laboratory for both of the Ba122 and  BaK122  crystals at different
temperatures from 40 to 180 K, with the
$c$-axis parallel to the external magnetic field ($H || c$). Several rectangular crystals  were stacked along  the
crystallographic
$c$-axis, and used for the NMR measurements corresponding to 10.6 mg  for BaK122 and 11.5 mg for Ba122. All
crystals were grown at Ames Laboratory and their magnetization measurements have been previously
reported~\cite{ni2008}. The NMR spectra were collected by a field sweep method, except for  the spectra at 6.4 T, which
were collected by frequency sweep. The temperature variation of the  spectra for BaK122 was measured at 6.4 T and
62.21 MHz but only at higher temperatures with the larger fields,  10 to 14 T.  A typical
$\pi/2$ pulse length was 1.8 - 2.5
$\mu$s, depending on experimental conditions, where we define this  pulse length as one that maximizes the echo
amplitude. The spin-lattice relaxation time ($T_1$) at 100 K at 8.5 T is
$\sim 17$ ms. There is a shift in the spectrum for the orientation
$H\perp c$, indicating an anisotropic Knight shift, which we will not  dwell upon here.  A similar kind of anisotropy
in the Knight shift has also been reported by Baek {\it et  al.}~\cite{baek2008} for a Ba122 crystal grown by the Sn
flux method.

\section{Impurity Effects on NMR}

NMR measurements give both static and dynamic information about the local magnetic field and the electric field gradient
distributions associated with impurities.  The NMR spectrum can be shifted by local magnetic fields from an impurity
magnetic moment, either through direct interactions via dipole-dipole coupling between local moments and the nucleus,
or through indirect interactions involving hyperfine coupling to conduction electrons (indirect exchange) or orbital
electrons (super exchange).  The indirect interaction has an oscillatory character (see the review by
Alloul~\cite{All74} for example), which for dilute concentrations of impurities can be recognized from the magnitude of
the effective field that they generate, defined later in this paper.  Since the impurity effects associated with the
indirect interaction are largely spatially inhomogeneous and are averaged over an effective field that is oscillatory,
one typically finds significantly reduced paramagnetic shifts in the average local field displayed in the Knight
shift~\cite{Alloul2009,alloul1991,bobroff1997,bobroff1999,ishida1996}.  On the other hand the distribution of these
shifts given by the NMR linewidth can be substantial.  It is also well established that paramagnetic
fluctuations from impurities can ``wipe-out'' contributions to the NMR from nearby spectator nuclei owing to
short spin-spin relaxation and/or short spin-lattice relaxation times.  

\section{Results and Discussion}

\begin{figure}[!ht]
\centerline{\includegraphics[width=0.65\textwidth]{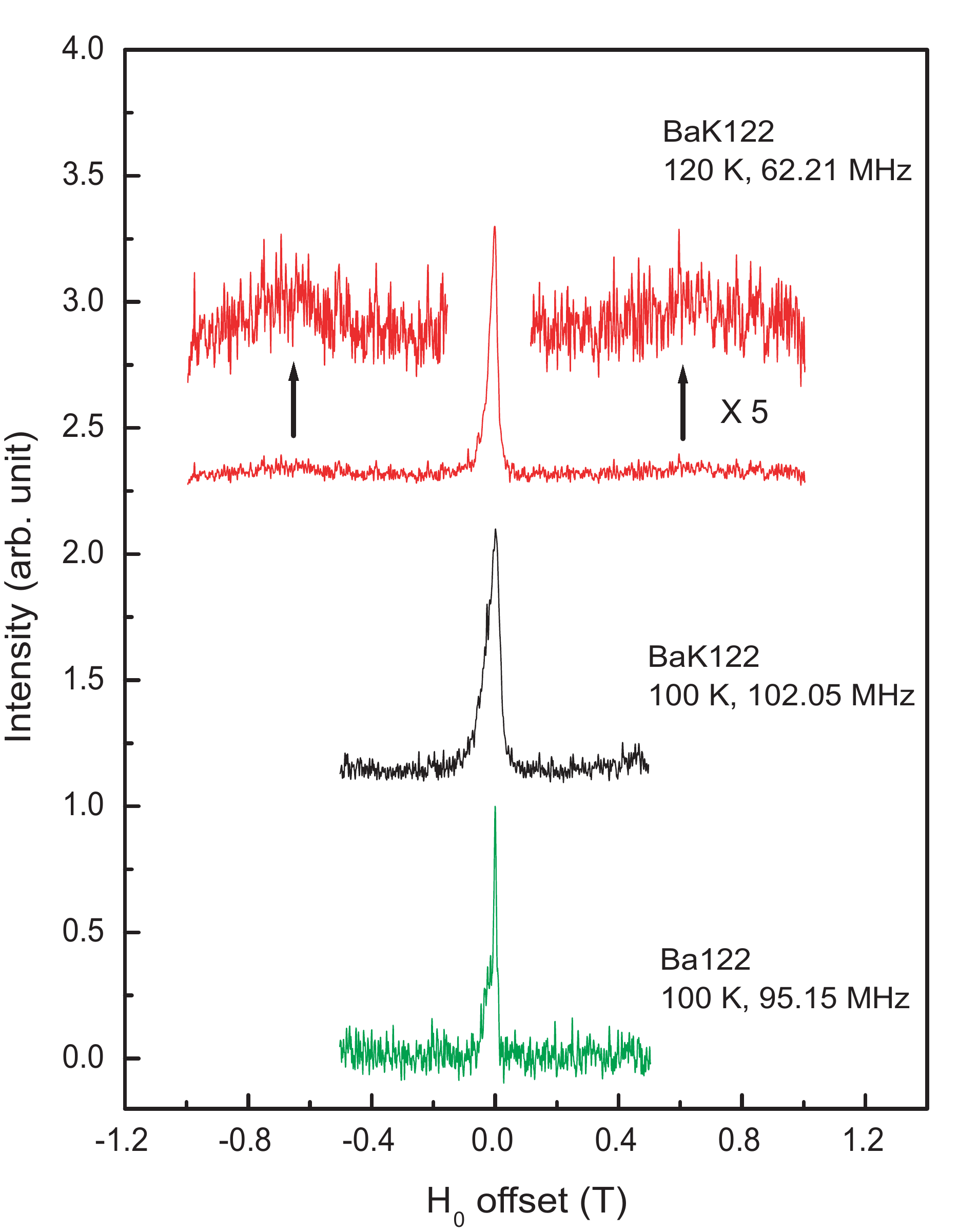}}
\caption {$^{75}$As NMR spectra for the Ba122 and BaK122 crystals,  obtained by field sweep. The uppermost trace is
magnified by a factor of five to  show the  quadrupolar satellite transitions.}
\label{fig1}
\end{figure}

\begin{figure}[!ht]
\centerline{\includegraphics[width=0.75\textwidth]{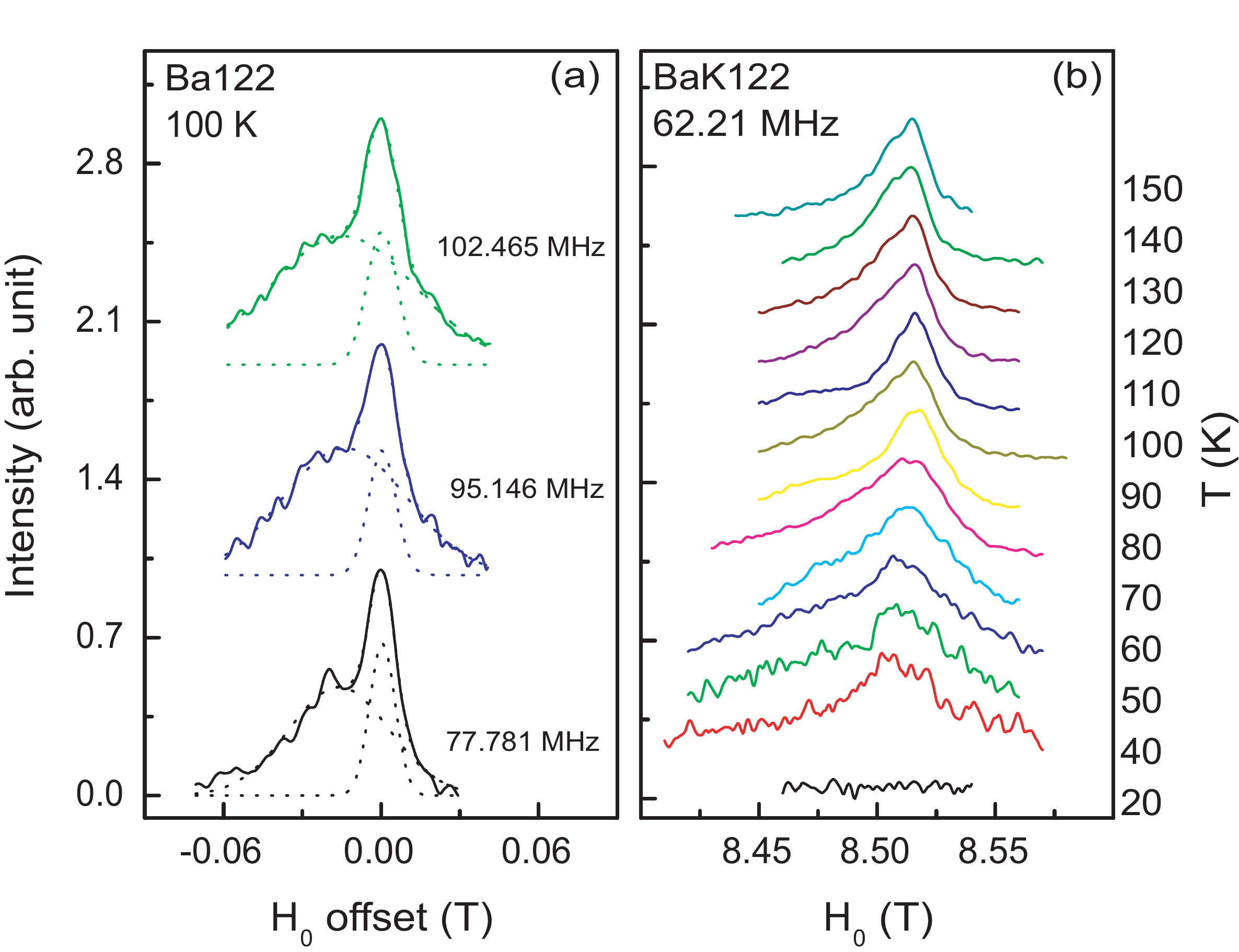}}
\caption{Field sweep $^{75}$As NMR spectra (a) for the Ba122 at  different Larmor frequencies at 100 K and (b) for
BaK122 at different temperatures for the Larmor frequency  62.21 MHz. For both  samples the line broadens with
increasing field and decreasing temperature. The signal  intensity decreases dramatically below 80 K, and eventually
become unobservable around 30 K (see  Fig.~3). In panel (a) the dotted lines indicate a decomposition of the spectra
into broad and narrow components. }
\label{fig2}
\end{figure}

  Fig.~\ref{fig1} shows the full spectra for
$H || c$ and in more detail in Fig.~\ref{fig2}, a representative  selection from our data.  We found the spacing 
between the satellite transition and the central transition,
$\nu_{Q}$,  to be  $\sim$ 5  MHz.

We first discuss the parent Ba122 crystal measurements at  temperatures above the structural transition comparing our
results with those of Refs.~\cite{baek2008} and 
\cite{kitagawa2008}. Fig.~\ref{fig2}a shows the evolution of the spectra with magnetic  field for Ba122. Our
spectra have a double peak structure, more prominent above 100 K,  also observed by Baek
\textit{et al.}~\cite{baek2008}. However, the full-width-at-half-maximum (FWHM) of our broad peak  is almost 4 times
that of Ref.~\cite{baek2008} (see Fig.~\ref{fig4}).  A  weaker temperature dependent
magnetization was reported by Baek {\it et al.}~\cite{baek2008}  compared to the present case~\cite{ni2008},
consistent with their narrower NMR line.  However, the spectra reported by Kitagawa
\textit{et al.}~\cite{kitagawa2008} are much
narrower for their self-flux grown crystals as compared to Sn-flux grown crystals, and at 140 K consist of a single  narrow line, having a FWHM of
3.5 kHz.  The second  order quadrupolar broadening of the central  transition should be zero for the $H || c$
orientation, assuming that the asymmetry parameter
$\eta$ = 0.   The central transition spectra for both doped and  undoped crystals are too broad to be accounted for by
misalignment. The Sn-flux grown  samples are  believed to have Sn impurities (of order 1\%) incorporated in the crystal
structure which are responsible for  its paramagnetism~\cite{ni2008}. Kitagawa
\textit{et al.}~\cite{kitagawa2008} have reported a significant increase  in the linewidth for both the central and
satellite transitions in a Sn-flux grown Ba122 sample  having 1.5\% Sn. They attributed this to disorder caused by Sn
impurities. Baek {\it et  al.}~\cite{baek2008} suggested that the broad line arises from an impurity phase,
containing As, which they estimate to be about 40\% of the total. The narrow line, or the `primary' peak in their
report, showed the existence of significant quadrupolar broadening (the linewidth decreases with increasing field).
We do not believe that there is a magnetic `second' phase.  Rather all of our data have substantial magnetic
broadening. This indicates, that there is a wide magnetic field
distribution that persists to the nanometer scale, typical of  magnetic interactions with the nuclear spin.

\begin{figure}
\centerline{\includegraphics[width=0.75\textwidth]{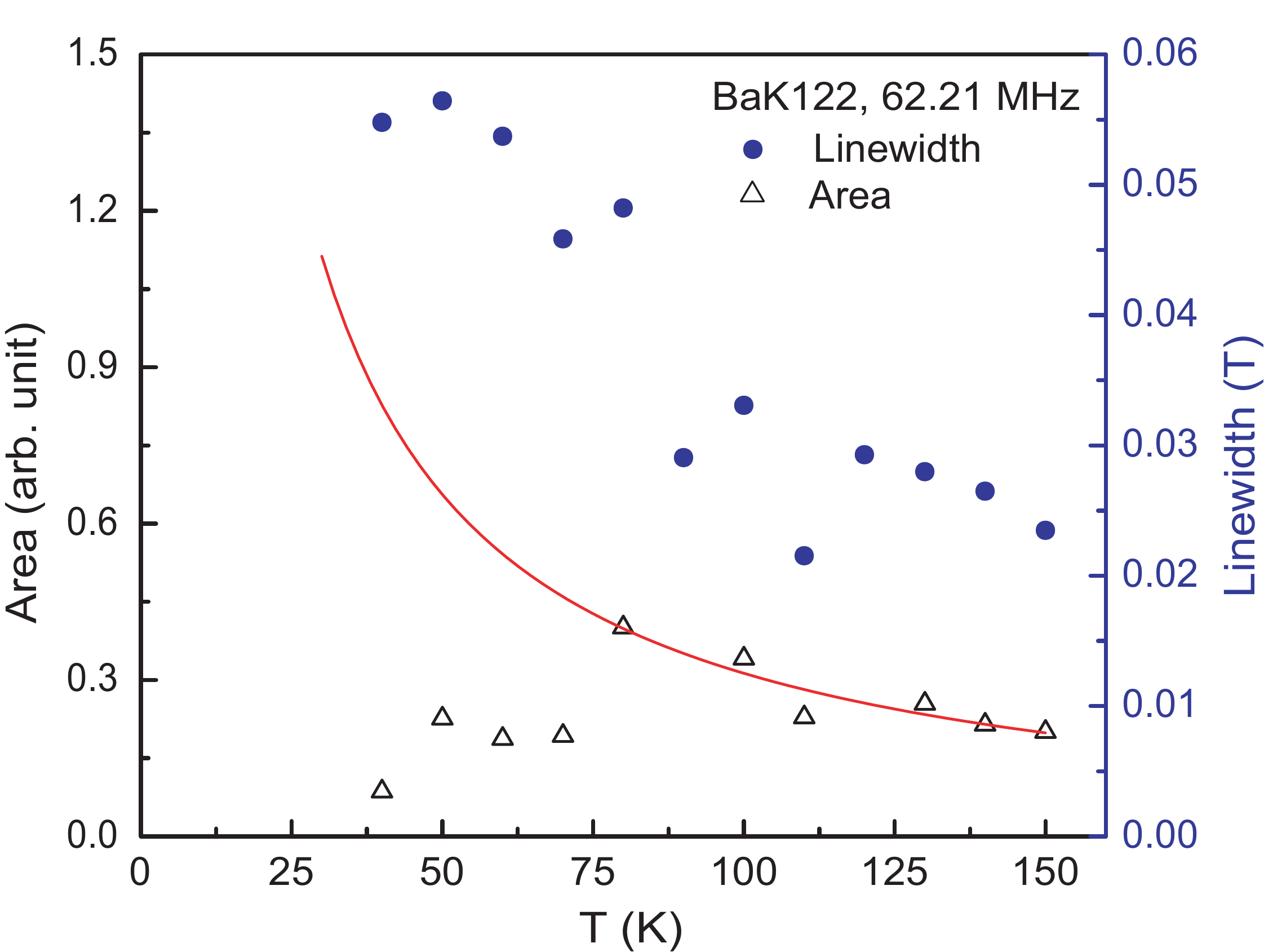}}
\caption{Temperature dependence of $^{75}$As linewidth and the area (integral) of the spectra for BaK122 at the Larmor
frequency 62.21 MHz. The solid line is the nuclear-Curie law  fit expected for the NMR signal amplitude. We have also
observed this deviation of the area of the  spectra from the Curie law below 80 K for Ba122, a consequence of extreme
magnetic NMR broadening.}
\label{fig3}
\end{figure}

\begin{figure}
\centerline{\includegraphics[width=0.75\textwidth]{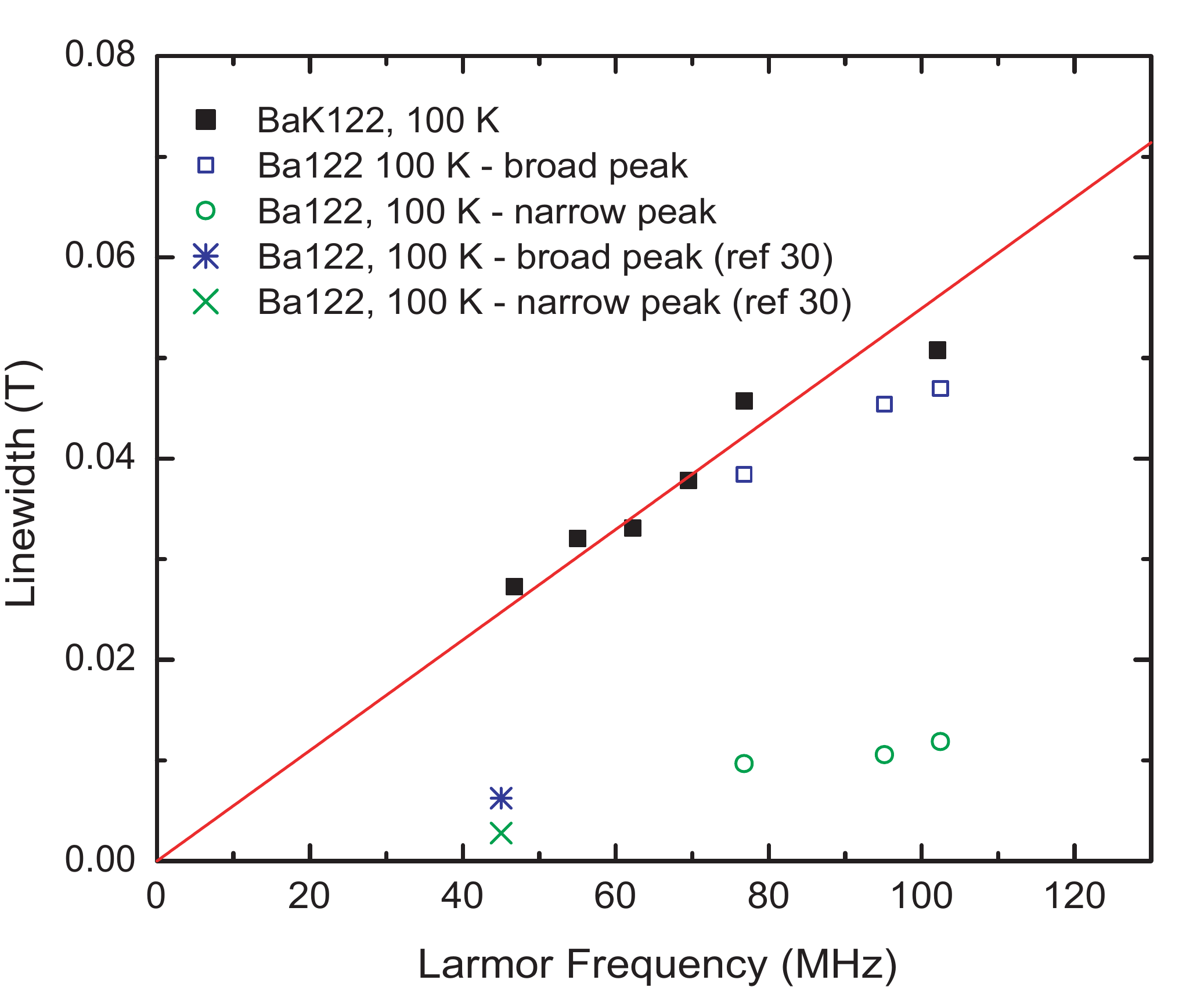}}
\caption{Field dependence of $^{75}$As linewidth for BaK122 and Ba122  at 100 K. For Ba122 the spectra could be
deconvolved into narrow and broad peaks; (see Fig.~2). Data  from Baek {\it et al.}~\cite{baek2008}  for the Ba122
compound is also shown.}
\label{fig4}
\end{figure}

In order to investigate the origin of this broadening we have studied  BaK122 crystals where magnetic ordering is
believed to be suppressed by doping with potassium.  The  spectra are broader than for Ba122, and cannot be resolved
into broad and narrow components, as was the  case of Ba122.  We have explored evolution of the spectra as a function
of both temperature and  magnetic field (a representative plot of the temperature evolution of the spectra at 62.21 MHz
is shown in  Fig.~\ref{fig2}b) from which we conclude that the NMR line broadening is a result of static magnetic 
field  distributions that become progressively more severe at lower temperatures, rendering the NMR  line unobservable
in the superconducting state. This is evident in Fig~\ref{fig3}, which shows  the variation of the linewidth (FWHM),
and the total nuclear  magnetization (proportional to the number of resonating nuclei, given by the area of the
spectrum) as a  function of temperature at 62.21 MHz. The total nuclear magnetization deviates from the 
nuclear-Curie law below 80 K indicating that there is additional line broadening too large for us to fully observe
the  resonance. However, it is clear that the NMR linewidth, which is to say the magnetic field  distribution,
increases with decreasing temperature and increasing applied magnetic field.  We note that
  the total nuclear magnetization for Ba122 also deviates from the  nuclear-Curie law below 80 K in the SDW state.

The linear variation of the linewidth of the BaK122 spectra with  magnetic field, measured at 100 K (Fig.~\ref{fig4}),
indicates that the linewidth is predominantly of  magnetic origin. Fig.~\ref{fig4} also shows the dependence on
magnetic field of the linewidth of the broad  and narrow components of the spectra for Ba122. The  field dependence of
the linewidth of the broad  component is similar to that of BaK122, {\it i.e.} proportional to the magnetic field
indicating that the  origin of the line broadening is the same for both samples, and in particular for BaK122, this
magnetic  inhomogeneity coexists with the superconducting state since the NMR line remains too broad to observe  below
$T_c$.

We express the total linewidth phenomenologically as,
  \begin{equation}
  \Delta \nu = \Delta \nu_{\mathrm{imp}} + \Delta \nu_ {\mathrm{intrinsic}},
  \label{eqn1}
  \end{equation} where $\Delta \nu_ {\mathrm{intrinsic}}$ is the  intrinsic linewidth of the spectrum in the absence of
magnetic inhomogeneity and for
$^{75}$As, this would mostly be due to quadrupolar broadening. The  fact that the linear fit in Fig.~\ref{fig4} passes
through zero indicates that the quadrupolar  contribution to the linewidth is negligible compared to the effect of
magnetic impurities. In  contrast, self-flux grown Ba122 crystals have a narrow NMR spectrum~\cite{kitagawa2008} and
an absence of  paramagnetic impurities.  Based on the results in Fig.~4 for BaK122, we expect that $\Delta
\nu$ should be proportional to the bulk magnetization~\cite{ni2008}, $M(T,H_0)$, 
\begin{equation}
\Delta \nu \propto M(T,H_0) \propto \chi(T) H_0.
\label{eqn2}
\end{equation}

\begin{figure}[!ht]
\centerline{\includegraphics[width=0.75\textwidth]{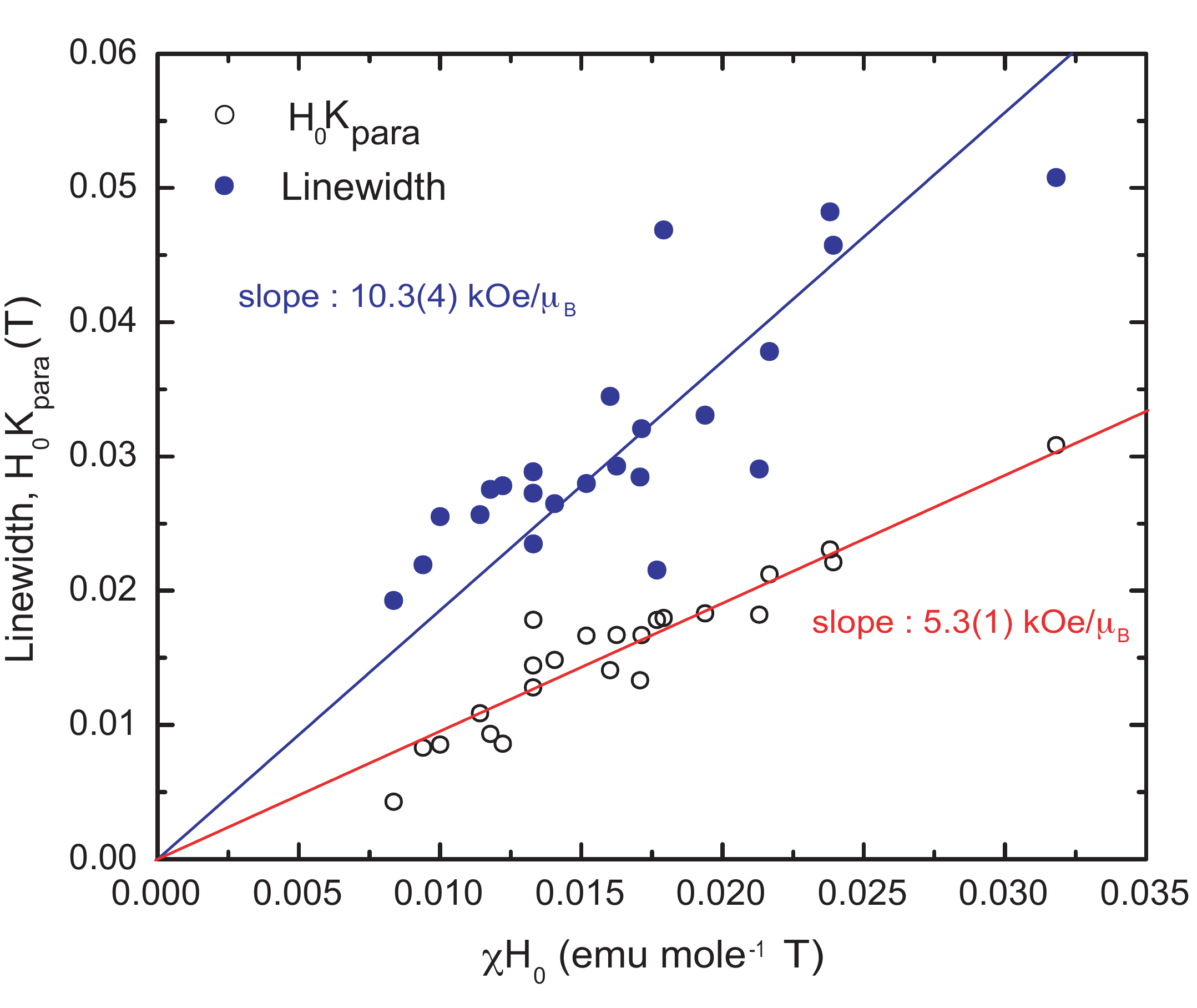}}
\caption{Combined plot of linewidth (FWHM) and $K_{\mathrm{para}} \times  H_0$ versus $\chi H_0$. The linear fit
indicates that, both the linewidth and $K_{\mathrm{para}}$ have the same field and temperature dependence as does the
magnetization~\cite{ni2008} which follows a Curie-Weiss law in the range of our NMR experiments.}
\label{fig5}
\end{figure}

As Fig.~\ref{fig5} shows, the linewidth is indeed proportional to $\chi(T)  H_0$. A  Curie-Weiss type temperature
dependence of
$\Delta
\nu_{\mathrm{imp}}$ is well known in the case of cation doped  YBCO~\cite{alloul1991, bobroff1997, bobroff1999} in
$^{17}$O, $^{79}$Y, and $^{7}$Li NMR and is  also observed for $^{17}$O NMR in pure BSCCO crystals where the
magnetic impurity has been associated with  the oxygen dopant~\cite{Che08}.

The Knight shift, $K$, calculated from the first moment of the spectrum relative to the bare nucleus, can be expressed as,
\begin{equation} 
K = K_{\mathrm{para}}+ K_s+K_{\mathrm{orb}},
\label{eqn3}
\end{equation} 
where, $K_s$ and $K_{\mathrm{orb}}$ are the spin part and the orbital part of the Knight shift,
respectively, and expected to be temperature independent in the normal state. 
$K_{\mathrm{para}}$ is the shift due to magnetic impurities, and can be written as,
\begin{equation} K_{\mathrm{para}} = \frac{1}{\mu_B}\chi(T)H_{eff}.
\label{eqn4}
\end{equation} Here, $H_{eff}$ is the effective magnetic field at the  As nucleus.

The data from all of our measurements have been plotted in  Fig.~\ref{fig5}, in the form
$K_{\mathrm{para}}H_0$   versus
$\chi H_0$.  From the linear fit to the total Knight shift at 6.4 T and 62.21 MHz we find $K_s + K_{\mathrm{orb}}$ to
be 0.37\%, somewhat  larger than for the undoped compound~\cite{kitagawa2008}, 0.3\%. The effective field we  obtain
from our fit is 5.3(1) kOe/$\mu_B$.  If the  magnetization from impurities were uniformly  distributed then the
effective field arising from direct dipole coupling of the impurity moment to the
$^{75}$As nuclei, based on
$B =
\mu_{0}(H + 4\pi M)$, would be 0.6 kOe/$\mu_B$, an order of magnitude  smaller than what we observe for the NMR shift
shown in Fig. 5.  Consequently, the effective field  must be enhanced by hyperfine coupling and leads us to the
conclusion that the As nuclei are indirectly  coupled to the magnetic impurities through the conduction electrons.

We cannot independently identify the magnetic contribution to depairing and  determine its role in the suppression of
$T_c$.  However, we note that the resistivity in the normal state  just above $T_c$ for BaK122 with
$\approx 1$\% Sn impurity~\cite{ni2008} is larger than for the pure compound~\cite{rotter2008} by $\sim 0.25$
m$\Omega$-cm, almost a factor of three; an  impurity effect which
  is significantly larger than for 1\% Zn substituted for Cu  in YBCO$_7$~\cite{Chi91}. This suggests that the
scattering mechanism for Sn in BaK122 might likely  be identified with its comparatively larger magnetization and
therefore be principally magnetic. So then  it is puzzling why the transition temperature in BaK122 is only suppressed
by $\sim 20$\%.  One  possible explanation might be that the magnetic impurity distribution, although very broad, might
be strongly  correlated, which is to say that there are regions with fewer impurities on the coherence length scale 
that dominate the overall transition temperature via the proximity effect.  This phenomenon is  well
established~\cite{Hal08} for impurities in superfluid $^3$He where it is shown that the transition temperature is
suppressed much less than the amplitude of the order parameter for a wide range in the strength of pair breaking.

\begin{figure}[!ht]
\centerline{\includegraphics[width=0.75\textwidth]{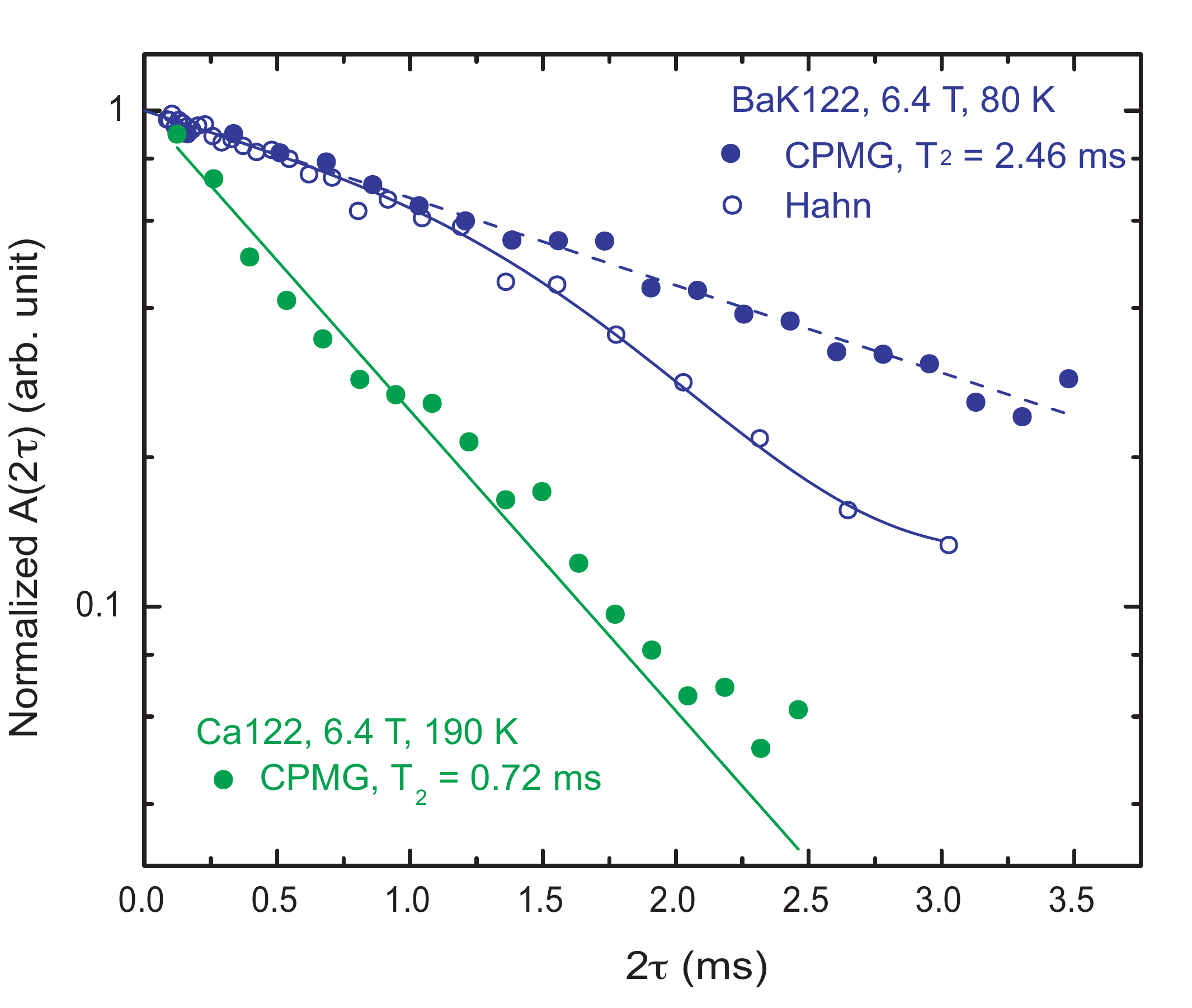}}
\caption{Spin-spin relaxation profile. The decay profile of  transverse magnetization. The time evolution of the echo
amplitude after a standard Hahn pulse sequence comparing  BaK122 and Ca122, and after a Carr-Purcell-Meiboom-Gill
sequence (CPMG) for BaK122.  }
\label{fig6}
\end{figure}

Also in Fig.~\ref{fig5} we show the linewidth from the  paramagnetically broadened NMR spectra as a function of
$\chi H_0$.  The effect of impurities is to produce a  strongly inhomogeneous field distribution throughout the
sample.  In fact, as a point of  comparison, consider the extreme model of a very inhomogeneous field distribution
taken to be constant, up to a  cutoff.  In this case the width of the distribution would be exactly half of the
average.  This  situation is rather similar to our spectra for BaK122, as shown in Fig.~5, where the linewidth is
twice the  Knight shift. 
From this perspective it is evident that the magnetic field distribution is indeed very broad.  Nonetheless, the
absence of a narrow component in the spectrum indicates that the field distribution  persists to the nanometer scale,
typical of magnetic interactions with the nuclear spin.

All of our NMR experiments are based on the formation of a spin echo  after two pulses, mostly involving the Fourier
transform (FFT) of the echo. Ideally the pulses are a 
$\pi /2 - \tau - \pi - \tau -$ echo sequence (Hahn echo). During the time $2\tau$, development of spin  decoherence
reduces the echo amplitude and, in principle, can alter the shape of the FFT spectrum.  Decoherence is determined by
spin-spin relaxation for which an example is shown in  Fig.~6. Here we compare  the relaxation profiles for the
crystals of BaK122 and Ca122, above the temperature for the  structural transition for the latter. Additionally we show
the echo amplitude evolution in BaK122 using the  Carr-Purcell-Meiboom-Gill (CPMG) sequence
$(\pi /2)_x - \tau - \pi_y - \tau - $echo$ - \tau - \pi_y - . . .$  where the indices $x, y$ label RF pulses with
orthogonal phases (see for example, Ref.~35).

For a static, homogeneous, environment of the nuclear spin its  spin-spin relaxation time, $T_2$, should be less than
that determined by the nuclear dipole-dipole coupling. 
 From the crystal structure we calculate that
  $T_2$ from this coupling is 0.72 ms for BaK122, and 0.59 ms for Ca122, which has negligible impurities. The calculated $T_2$ is close to what we observe in the case  of the Ca122 crystal.  However, for BaK122 the recovery profile  gives a larger $T_2$ than expected by
$\approx$ 2.5, for small $2\tau$ values. This result indicates that there is quenching of the  dipole-dipole
interaction for BaK122, presumably from magnetic field gradients producing local field  differences greater than the
dipolar field of the near-neighbor As atoms.  We were then led
to explore if these  field gradients were dynamic by application of a CPMG sequence that can reduce or eliminate
their  stochastic effect~\cite{Sig01}. Indeed, the CPMG recovery profile develops even less rapidly  than  that for
the Hahn echo suggestive of the existence of these field fluctuations.  Theoretical  analysis~\cite{Sig01} of the Hahn
echo profile gives the fit shown in Fig.~6 for a characteristic field fluctuation  frequency ($\omega$) of just 2.2
kHz.  Although the fit is excellent, we should be careful about this interpretation since there is not a well-defined
$\pi-$pulse.  The RF-pulses are rather inhomogeneous over the sample  and, secondly,  the NMR line for BaK122 is very
broad, meaning that much of the received signal is off-resonance. In  this regard we note that the CPMG experiment on
Ca122, for which the crystal has a similar geometry and consequently the same inhomogeneous RF-field, has a CPMG
recovery profile which is the same as for the Hahn echo.  This is consistent with our expectation for a pure
compound in the absence of field fluctuations.  A similar effect was observed in
$^{195}$Pt NMR in nanometer size unsupported Pt particles~\cite{Yu1993}. There the Hahn-echo $T_2$ of the Pt nuclei
on the particle surface, is much shorter than the CPMG $T_2$, which can be accounted for by an inhomogeneous
magnetic field with field fluctuations arising from spin-diffusion. If the local fields are indeed fluctuating in
BaK122, in the kHz range, this is so low in frequency that it suggests that the magnetism is a collective phenomenon,
possibly super-paramagnetic, rather than being associated with independent fluctuations of local moments.  Most
importantly however, is that we have found an unusually long
$T_2$ which is a an indication of large local magnetic field gradients in BaK122.

We have also compared
spin-lattice relaxation in BaK122 and in Ca122.  For the former we find $T_1$ at 100 K $\sim 17$ ms,  in comparison
with 7.6 ms at  190 K for Ca122. The
$T_1$ for BaK122 is comparable to the $T_1$ in Ba122 reported in Ref~30 at the same temperature.
Despite the existence of magnetic impurities in BaK122, its relatively long $T_1$ gives no  indication for a
contribution  to spin-lattice relaxation from field fluctuations, possibly  because their frequency is too low
compared to the Larmor frequency. Lastly, it is essential in an NMR  experiment, to check how much of the sample is
contributing to the signal.  It is meaningless to interpret  NMR results if the vast majority of the sample, relevant
in comparisons with other experiments, does not contribute to  the NMR signal. We have made a direct estimate of the
NMR active fraction of the samples comparing BaK122  and Ca122 and have found that the NMR signal strength is
comparable, taking sample shape and size into account and using the same NMR conditions. Consequently, we believe that
the results we report are  representative of the entire sample with the exception, as noted above, for data below 80
K in BaK122 and Ba122.

\section{Summary}

The magnetization from impurities  in our BaK122 crystals, grown from  Sn flux, is exceptionally large. In fact the
magnetization is $\sim$ 60 times larger than that of  La$_{1.85}$Sr$_{0.15}$CuO$_4$ for which superconductivity is
completely suppressed by Al substitution.  The  large Curie-Weiss-like magnetization in BaK122 results in a significant
broadening and displacement of the
$^{75}$As NMR spectrum.  Since there is no narrow component we infer  that the magnetic inhomogeneity persists to a
microscopic scale.   The local moments associated with  these impurities appear to be coupled to the As nucleus by an
indirect interaction through the  conduction holes with an effective field H$_{eff}\sim 5$ kOe/$\mu_B$.  We suggest
that the impurities are  associated with Sn substituted on the As site, broadly distributed.  Superconductivity in
BaK122 takes place  in a backdrop of inhomogeneous magnetism associated with these impurities. The extent of pair 
breaking from magnetic impurity scattering can only be qualitatively linked to their paramagnetism, or possibly 
super-paramagnetism.  Nonetheless, our BaK122 crystals are indeed superconducting and exhibit a relative  suppression
of the transition temperature of only 20\%.  We attribute the robustness of  superconductivity in the presence of
magnetic impurities to strong spatial correlations between the impurities, a non-uniformity that is also apparent in
the large magnetic field distributions and field gradients which we observe  directly with NMR.

\ack

We thank N. Curro, V. Mitrovic\'{}, Y. Furukawa, J. Sauls and A. Chubukov for helpful  discussions.  We acknowledge
support from the Department of Energy, Basic Energy  Sciences under Contracts No. DE-FG02-05ER46248 (Northwestern
University)  and No.  DE-AC02-07CH11358 (Ames Laboratory). Work at high magnetic field (14 T) was performed at the
National High Magnetic  Field Laboratory with support from the National Science Foundation and the State of Florida.
One of the authors (M.L.) was supported by Kunkuk University in Seoul, South Korea.

\section*{References}

\end{document}